\documentclass[12pt]{article}

\usepackage{geometry}                
\usepackage{graphicx}
\usepackage{amssymb}
\usepackage{epstopdf}
\newcommand\0[1]{{\mathrm{#1}}}

\newcommand\2[1]{{\mathbf{#1}}}

\newcommand\6[1]{{\overline{#1}}} 

 \newcommand\8[1]{{\widehat{#1}}}

\newcommand\Alg{\mathop{{\0{Alg}}}\nolimits}
\newcommand\apost{\mbox{\bf '}\,}

\newcommand\Av{\mathop{{\0{Av}}}\nolimits}

\newcommand\miss{\mathrel{{\kern3pt\backslash\kern-8.1pt\bigcirc}}}
\newcommand\Oplus{\bigoplus}

\newcommand\Bar{{{}\vrule height 8pt width 1pt depth 3pt{\kern3pt}}}

\newcommand\Cliff{\mathop{{\0{Cliff}}}\nolimits}
\newcommand\Deg{\mathop{{\mathrm{Deg}}}\nolimits}

\newcommand\Dual{\mathop{{\mathrm {Dual}}}\nolimits}
\newcommand\Dup{\mathop{{\mathrm {Dup}}}\nolimits} 
\newcommand\fermi{\mathop{{\0{Fermi}}}\nolimits} 
\newcommand\Fermi{\fermi}

\newcommand\Exp{\mathop{{\mathrm{Exp}}}\nolimits} 
\newcommand\Grass{\mathop{{\mathrm{Grass}}}} 

\newcommand\Q{{\3Q}} 

\newcommand\Qi{{\8{\imath}}}
\newcommand\Rank{\mathop{{\mathrm{Rank}}}\nolimits}

\newcommand\slin{\mathop{{\mathrm {sl}}}\nolimits} 
\newcommand\SL{\mathop{{\mathrm {SL}}}\nolimits} 
\newcommand\so{\mathop{{\mathrm {so}}}\nolimits} 
\newcommand\SO{\mathop{{\mathrm {SO}}}\nolimits} 

\newcommand\Spin{\mathop{{\mathrm {Spin}}}\nolimits}

\newcommand\Z{{\vrule width.5pt height0pt depth0pt \vrule width 2pt height 2pt depth 0pt  \vrule width.5pt height0pt depth0pt}}
\newcommand\Y{\Z}

\newcommand\ox{\otimes}

\newcommand\BEQ{\begin{equation}}
\newcommand\EEQ{\end{equation}}
\newcommand\BEN{\begin{enumerate}}
\newcommand\EEN{\end{enumerate}}
\newcommand\BTA{\begin{table}}
\newcommand\ETA{\end{table}}
\newcommand\BIT{\begin{itemize}}
\newcommand\EIT{\end{itemize}}
\newcommand\BEA{\begin{eqnarray}}
\newcommand\EEA{\end{eqnarray}}

\newcommand\fro{\leftarrow}

\newcommand\cto{\kern3pt-\kern-9pt\succ}

\newcommand\cfro{\prec \kern-7pt -\kern3pt{}}

\newcommand\lar{\leftarrow}

\newcommand\rar{\rightarrow}

\newcommand\mapsfro{{\fro}\kern-4pt\rule{.5pt}{5pt}}

\newcommand\dar{{\downarrow}}
\newcommand\uar{\uparrow}

\newcommand\Diagr[8]
{\begin{array}{rcccl}%
&              &{\scriptstyle #5}           &                                            &\cr
& #1                                          & \rar& #2                                      &\cr
\cr
{\scriptstyle #8}&\dar \;\; &         &\;\;\dar&{\scriptstyle#6} \cr
\cr
& #4                                   & \,\rar \,& #3                                      &\cr
&                           &{\scriptstyle#7}&  			             &     
            \end{array}}

\newcommand\DDiagr[8]
{\begin{array}{rcccl}%
&              &{\scriptstyle #5}           &                                            &\cr
& #1                                          & \lar& #2                                      &\cr
\cr
{\scriptstyle #8}&\uar \;\; &         &\;\;\uar&{\scriptstyle#6} \cr
\cr
& #4                                   & \,\lar \,& #3                                      &\cr
&                           &{\scriptstyle#7}&  			             &     
            \end{array}}

\font\sm = cmr10 at 8 pt

\title{\bf Recursive quantum gauge theories}
\author{David Ritz Finkelstein\\
\small Georgia Institute of Technology, Atlanta, Georgia\footnote{Emeritus.}\\
\small finkelstein@gatech.edu}

 \date{}                                      

\begin{document}
\maketitle

\abstract{Quantum gauge theories with finite-dimensional representation spaces  are constructed  that can have canonical gauge field theories as singular limits. 
They describe nature as a recursive quantum assembly by iterating Fermi-Dirac quantification. 
Six iterations  are necessary and sufficient for present physics. 
The gauge structure, the spin-statistics correlation,
the space-time metric, and the Higgs field 
are modeled.

}

\section{Finite quantum theories} 

\label{S:REGULAR}
Canonical gauge theories like general relativity  and the Standard Model are triply singular, due to
(1) the canonical commutation relation between the  differentiator  $\partial_{\mu}$ and the coordinate
$x^{\mu}$,
(2) the canonical commutation relations between the gauge connection $\Gamma_{\mu}$ 
and its canonical-conjugate gauge field,
and (3)  the vanishing Hessian determinant of the gauge-invariant action
  with respect to time derivatives of the gauge vector potentials.
The current gauge theories are often optimistically supposed to be singular limits of some regular  theory to come.
A kinematics for such a regular theory is developed here
 to the point where it provides finite correspondents for the
 operators in the usual action principle for a general gauge group.
 Dirac often emphasized the need  for a finite theory;
 Bopp and Haag (1950) posed a regularity principle  for spin models,
 and Segal (1951) posed one for quantum field theory in general.
\nocite{BOPP1950}

The proposed quantum theory, termed recursive,  represents 
the system as a recursive quantum assembly.
Its modules have Fermi-Dirac statistics, and are modularizations,
or unitizations, of  like assemblies of a lower level, or rank.
Each assembly is also interpreted  as a quantum topological simplex with its constituent modules as its vertices.
Let us call such a recursive assembly or simplex a {\em plexus} for short.
Multiple quantification was proposed by Weizs\"acker (1955), John Baez (2005), 
and others.\nocite{WEIZSAECKER1955,BAEZ2005}

A recursive assembly has a natural gauge theory, discrete for classical systems, continuous for quantum ones.
For  Weyl the  gauge was a machinist's gauge block or a carpenter's gauge,  a metaphor for
a precise standard of measurement that  undergoes a
non-integrable change during its transport around a tube of magnetic flux.
The gauge block for gravity is a vector in a  Minkowski tangent space of the space-time manifold as a standard of direction.
In Dirac's revision of Weyl,  and in the Standard Model, the gauge block is the quantum particle under study, undergoing
a semi-simple group of its representation space 
when it is carried around flux of the gauge field.
The gauge block of a  quantum plexus
 is the module  of a certain stage of assembly. 
 Let us  call it the {\em cell}
of the plexus.
The gauge Lie algebra of the first kind describes the statistics of the vertices of the cell.
It induces transformations of the gauge Lie algebra of the second kind 
in assemblies representing events, perhaps two ranks higher than the cell.
Its non-integrability, as expressed by the gauge curvature, results
from dislocations in the organization of the plexus.

Stages of assembly 
are linearly ordered by rank (as in the von Neumann universe of sets).
For example, since dynamical variables are functions of time, 
they have higher rank than time.
In the kinematics considered here,
the cell  has  some rank $C\approx 4$, and
 the space-time event and field have a
rank $E\approx 6$, which is large enough
for present physics.


\subsection{\em Quantization as quantification}
\label{S:QQ}

It is well known that ``second quantization'' is a misnomer for a process
that does not quantize but quantifies,
passes from  one quantum  to an assembly of many.
Its commutation relations  express statistics, not just complementarity.

The same can be said of canonical quantization, however.
In general, canonical quantization can be broken down schematically into  two steps:
\BEN
\item Choose a vector space
of  canonically conjugate dynamical variables.
\item Form polynomials in the chosen variables subject to 
graded canonical commutation relations.
\EEN
In retrospect, the first step specifies the representation space of a certain sub-quantum, 
and the second step is a quantification, assembling replicas of that sub-quantum into
a quantum system with Bose-Einstein or Fermi-Dirac         
statistics.  

In recursive  quantification, the initial vector space is $\1R$, and the single step is quantification 
with Fermi-Dirac statistics,
iterated recursively as often as necessary.
Six recursions are necessary and perhaps sufficient for present physics.

Thus one functor $\Grass$ is used  whenever we
quantify, quantize,  or gauge,
 to convert a
 one quantum theory to a many-quantum one, and one gauge block to many.
 
 While ``second quantization" is a misnomer, ``second quantification" may be accurate.
It is unreasonable to assume that some quantum non-commutativity comes from statistics and some
from some more mysterious complementarity, with no experimental evidence for such a division.
Let us suppose here that all complementarity expresses statistics on some level.

\subsection{\em The laminar phase}
 \label{S:LAMINA}
 
The  Standard Model groups can all be represented faithfully
on about $16$ vertices.
If there are $\4N$ vertices in quantum history,
  a random simplex would have $\4N/2\gg 16$  vertices
 on the average.
 Clearly
the ambient plexus is not random but highly organized
as a lamina,   
in that the spectral multiplicity of its
 basic variables are either
$\le 10^1$.
or $\gg 10^{15}$.
Let us call them the short variables and the long.
Short variables may be further classified as longitudinal 
if they transform under transformations of the long variables, or else transverse.
The  space-time
coordinates $x^{\mu}$ and  momentum-energy coordinates $p_{\mu}$, called {\em orbital}, and gauge boson field variables, are long;
spins are longitudinal short variables; 
the  hypercharge $y$,                                                                                                                                                                                                                                                                                                                                                                                                                                                                                                                                                                                                                                                                                                                                                                                                                                                                                                                                                                                                                                                                                                                                                                                                                                                                                                                                                                                                                                                                                                                                                                                                                                                                                                                                                                                                                                                                                                                                                                                                                                                                                                                                                                                                                                                                                                                                                                                                                                                                                                                                                                                                                                                                                                                                                                                                                                                                                                                                                                                                                                                                                                                                                                                                                                                                                                                                                                                                                                                                                                                                                                                                                                                                                                                                                                                                                                                                                                                                                                                                                                                                                                                                                                                                                                                                                                                                                                                                                                                                                                                                                                                                                                                                                                                                                                                                                                                                                                                                                                                                                                                                                                                                                                                                                                                                                                                                                                                                                                                                                                                                                                                                                                                                                                                                                                                                                                                                                                                                                                                                                                                                                                                                                                                                                                                                                                                                                                                                                                                                                    
isospin $\tau^k$,    
color charges $\chi^c$,
and generation $\Gamma$ are transverse short  variables.
The  Higgs field at each event is long.

Let us provide a regular kinematics for the laminar variables;
that is,  one based on regular Lie algebras,
which are those with regular Killing form.
A regular theory must be quantum,
since commutative Lie algebras are singular,
and must have a finite-dimensional representation space for its operators.
like a finite aggregate of spins.
Let us suppose that their laminar structure originates much as a graphene does,
from organization of the plexus out of cellular elements.

A regular Lie algebra is semi-simple, a direct sum of simple ones.
But in a quantum theory such a direct sum represents a disjunction of possible theories,
 one of which
can be selected by a single measurement.
Let us therefore specialize from  regular Lie algebras to simple
with no loss of generality.

\subsection{Quantum space-times}
\label{S:QSPACETIMES}

In order to bring the spirit of quantum physics to bear on space-time,
we must reconsider the experimental  meaning of the event construct.
Einstein represented an  event
by a smallest possible occurrence,
such as
 a collision of two small bodies.
 We may accept this.
But then he took as its defining variables
only space-time coordinates  defined originally by a lattice of clocks
 and later by light signals.
Today this would appear as a gratuitous inconsistency, since it neglects the
influence of observation on the observed.
  If an event is a collision then 
the {\sc io} channels of the collision already  describe it.
 If these do not carry photons, they will carry
 other kinds of quantum that are just as good.
 A supernumerary lattice or light signal would participate in the event,
 resulting in a quite different event.
 
The insistence on space-time coordinates seems to
be a vestige of  the  Cartesian hypothesis that position in space
 and time are the necessary and sufficient  variables for a true model.
This was already displaced by Newtonian mechanics with its forces and masses.
The test of a quantum theory is not whether it permits a space-time picture 
but whether it represents experimental processes with their
observed spectra, 
 transition probabilities, and assembly relations,
including our macroscopic physical experience as a correspondence limit of large quantum numbers.
For this we require all the coordinates of \S\ref{S:LAMINA} at least initially..

\subsection{\em Spinorial quantum spaces} 

Feynman, Yang, and Penrose  made helpful attempts at quantizing the orbital  space-time variables.
Let us compare them in
natural units. 
Here   $\8x$ designates a quantized $x$;  $k\in \{1,2,3\}$;  $m\in \{1,2,3,4\}$; and
$\delta x$ indicates an ``atom" of $x$: a finite-difference element of $x$, to be summed later. The elements of these Lie algebras are dynamical transformations, some of which are symmetries as well.
 \BEQ
 \label{E:QSPACETIMES}
 \begin{array}{lllllll}
\mbox{Feynman (1941)}&  \delta \8x^m &\sim& \gamma^m\/,&\cr
\mbox{Yang (1947)}&\8x^m&\sim& i\eta^{[5}\partial^{m]},
& {\8p_m}&\sim& i\eta_{[6}\partial_{m]}. \cr
\mbox{Penrose (1971)}&  \delta \8x^k&\sim&\sigma^k, &  \cr
\mbox{Present}&  \delta \8 x^m&\sim& \gamma^{m5}, &
\delta \8 p_m&\sim& \gamma_{m6}\/.
\end{array}
\EEQ
 \nocite{FEYNMAN1941}
 \nocite{YANG1947}
 \nocite{PENROSE1971}
The Penrose (1971) and Feynman (1941)  quantum spaces still 
assume an absolute space or space-time points,
with coordinates $x^{\mu}$ but not $p_{\mu}$.
Guided by group simplicity, Yang (1947)  relativized space-time within a larger 
quantum phase space, {\em Yang space}, 
whose quantum coordinates include positions, momenta, 
boosts, angular momenta,
 and a quantized imaginary $L_i=\Qi$
 that we will interpret as electric generator and Higgs field;
 all on an equal footing.
 A quantum  space today
 must fit in  all the variables of \S \ref{S:LAMINA}
and their commutation relations.
This can be done in several ways for the transverse variables, corresponding to possible GUT theories,
so let us consider only the orbital variables at first.

Yang, following Hartland Snyder (1947),
 represented his algebra by differential operators on an infinite-dimensional function space, as
shown in (\ref{E:QSPACETIMES}).\nocite{SNYDER1947}
Let us retain the Yang Lie algebra for its regularity but replace the classical
$\1R^6$ infrastructure for its singularity,
in favor of
a quantum space that is regular  like Feynman's and Penrose's, 
with odd statistics  to go with its odd spin parity.
This is the bottom line of (\ref{E:QSPACETIMES}).
This finite-dimensional representation space must have an
indefinite metric.
The unit cell of a lamina may have the Yang dynamical group, but not the lamina itself,
whose organization
rearranges the Yang group
in the sense of  Umezawa (1993).\nocite{UMEZAWA1993}
A   quantum space-time  lamina of Connes (1994) has the singular long dimensions of a manifold
and regular short dimensions; here both are regular.\nocite{CONNES1994}

\section{The recursive Grassmann algebra}
 Fermi-Dirac quantification preserves regularity; 
 Bose-Einstein does not.
Let us use Fermi-Dirac statistics for each stage of  recursive assembly, therefore.

The
classical prototype recursive assembly
is  the {\em recursively finite set}.
The space $\3C$ of recursively finite sets is
made from ${}_0\3C:= \{\emptyset\}$
in a finite number
of stages of power-set formation ${}_{r+1}\3C=2^{({}_r\3C)}=: \exp{}_r\! \3C$;
prefixes are ranks.

We form the power set $2^{\3X}$ by  unitizing the sets in $\3X$, forming all possible disjoint unions of the resulting unit sets, and collecting them into $2^{\3C}$.
 all their disjoint unions.  
 Let us represent the disjoint union
as a multiplication operation $\vee$, often left unwritten,
and unitization
as Peano's monadic operation $\iota: x\mapsto \{x\}$.
Multiplication assembles modules, unitization modularizes assemblies.
Let us also interpret any recursive finite set as a recursive simplicial complex, 
or plexus, whose vertices are unit sets, monads.
The classical algebra $\3C$ of recursively  finite sets
with the operations $\emptyset,  \iota, \vee$
is  a  recursive binary exterior algebra (over itself). 
Its identity is $1=\emptyset$.
Every other element has square 0.

\subsection{\em Recursive quantification}
Let us use $\3C$  as armature on which to construct
an analogous recursive quantum algebra $\3Q$. 
The necessary minimum operations for  quantum recursive assembly
seem to be $1,  \iota,  \vee$,
and  $+$\/.
$\vee$ is now a Grassmann product, sometimes unwritten, used to
assemble quantum modules.
$\iota$ is now an upgrade of Peano's $\iota$  to 
 a linear operator that
 maps $\3Q$ onto its first-degree sector $\3Q^1\subset \3Q$,
by identification modulo the equivalences
\BEQ
\forall a,a\in\1R,\; Q,Q'\in \3Q: \iota(a'Q'+aQ)\equiv a'\,\iota Q' + a\,\iota Q\/.
\EEQ
Addition $+$ in $\3Q$ is quantum superposition.
These operations  are commonplace in quantum physics, 
though $\iota$ is usually tacit.
When we attach indices
to a scalar field to build a tensor field, we implicitly form a tensor product and then
 unitize it so that its factors are not confused with those of other similar tensors in a
product. 

$\3Q$ is a recursive Grassmann algebra, over itself;
$\3C$ corresponds to the rays of a classical basis for $\3Q$,
made without quantum superposition.

There is a standard way to propagate a metric from a 
one-quantum representation space
$\3V$ to the many-quantum space $\22^{\3V}:=\Grass \3V$.
Applied recursively to the natural metric $\|r\|=r^2$ on $\1R={}_0\3Q$, 
which is
stage 0 of $\3Q$,
it defines a Hilbert metric $\4H: \3Q\to \Dual \3Q$
that will serve as a reference metric on the representation space,
and a Hilbert  norm $\|Q\|_{\4H} :=\4H Q \circ Q$\/.

Also useful is the neutral  {\em duplex norm} $\4D$ on 
the {\em duplex space} $\Dup \3Q:=\3Q\oplus \Dual\3Q$,
whose value $\|Q\oplus Q'\|_{\4D}\|$
for any $Q\in \3Q, Q'\in \Dual \3Q$ is the valuation $Q'\circ Q$\/.
 The duplex space  occurs in the spinor constructions of Cartan (1913) and Chevalley (1954)
 and in the theory of Saller (2006), who calls it
the {\em quantum space} of the system.\nocite{SALLER2006a}

Chevalley (1954) and  Bohm (1962)  represented one-level simplicial complexes in the  Grassmann algebra on its vertices.\nocite{BOHM1962,CHEVALLEY1955}
The plexus iterates the functor $\Grass$ recursively.

%
%

To bring out the intended parallel to the classical power set $2^X$, let us also write
 the Grassmann algebra as
\BEQ
\Grass \3V =: \22^{\3V}\/.
\EEQ
The boldfaced $\22$ stands for the quantum binary decision, to be or not to be in a product,
which has to be made for each basis element in $\3V$,
and their quantum superpositions.
The many-quantum representation space
 is the binary exponential of the one-quantum
space as the many-body phase space is the binary exponential of the one-body.
Conversely, we recognize a system as composite if its representation space $\3X$ has a useful logarithm $\3V=\log_{\22} \3X$, in the sense that $\3X=\22^{\3V}$.

Let us also use the standard apostrophe notation
\BEQ
f\apost X := \{f(x)|x \in X\}
\EEQ
for ``the $f$'s of the $X$'s"\/.
Then
$\3Q=\Grass \3Q$ is
recursively generated from  
$\1R$
by iterating $\Grass \iota\apost$:
\BEA
{}_0\3Q &:=&\1R.\\
{} _{r+1}\3Q &:= &\Grass {}\!_{\1R}\,\iota\apost (_{r}\!\3Q) \sim  \Grass {}_r\!\3Q\/.
\EEA

$\3Q$ is doubly graded, by  polynomial {\em degree} $g$  and by 
von Neumann {\em rank} $r$:
\BEQ
\3Q=\Oplus_{g,r} \; {}^r\!\3Q^g\/.
\EEQ
Degree 
counts monadic factors $\iota Q$.
Rank counts nested $\iota$'s.
A polyadic in $\3Q$ that is homogeneous of degree $g$ is called a $g$-adic.

%
The {\em serial number} $q$ for the classical basis  element $e_q$  gives the order of  creation, when first priority of operation is given to 
1, second to $\vee$, and third to $\iota$,
and factors in a product are ordered by their serial number.
Then
\BEQ
\forall q,q'\in \1N:\quad e_{(2^q)}=\iota e_q,\quad e_{q+q'}=e_{q'}\vee e_q\/ 
\EEQ

Table \ref{T:POLYADICS} lists some basic polyads   $\3C\subset \3Q$,
 by stratum,
with their serial number and statistics (exchange parity).
 The basis tabulated is generated from 1 by $\iota$ and $\vee$,
 without quantum superposition,
 and so is called a classical basis.
 Its symbols may be regarded as  {\em hyperbinary numbers},
 two-dimensional positional notations 
 for their serial numbers $q=\sum q_n \Exp n$.
The positional values $\Exp n $ grow hyperexponentially with position $n$ where binary values $2^n$ merely grow exponentially. 
The hyperbinary coefficients $q_n$
have  growing ranges $0\le q_n < \Exp n$ 
instead of the fixed range $q_n=0,1$  of binary 
coefficients.

Stratum 4 would overflow the page;  Table
\ref{T:MONADICS} lists its monadics.
Stratum 5 in Planck-size characters would fill  the known universe many times.
\BTA [t]
\caption{Polyadics  and their statistics,   by rank and serial number}
\label{T:POLYADICS}
{
\[
\begin{tabular}{|c|ccccccccccccc|}
 \hline
  &$-$&$+$&$+$&$-$&$+$&$-$&$-$&+&+&$-$&$-$&$+$&$\dots$\cr
${\mbox 6}$ 
&
$\stackrel{\6{\6{\6{\6{\6{\6{\Z}}}}}}}{}$&$
\stackrel {\6{\6{\6{\6{\6{\6{\Z}}}}}}\,\6{\Z}}{}$&$
\stackrel {\6{\6{\6{\6{\6{\6{\Z}}}}}}\,\6{\6{\Z}}}{}$&$
 \stackrel {\6{\6{\6{\6{\6{\6{\Z}}}}}}\,\6{\6{\Z}}\,\6{\Z}}{}$&$
\stackrel {\6{\6{\6{\6{\6{\6{\Z}}}}}}\,\6{\6{\6{\Z}}}}{}$&$
\stackrel {\6{\6{\6{\6{\6{\6{\Z}}}}}}\,\6{\6{\6{\Z}}}\,\6{\Z}}{}$&$
\stackrel {\6{\6{\6{\6{\6{\6{\Z}}}}}}\,\6{\6{\6{\Z}}}\,\6{\6{\Z}}}{}$&$
\stackrel {\6{\6{\6{\6{\6{\6{\Z}}}}}}\,\6{\6{\6{\Z}}}\,\6{\6{\Z}}\,\6{\Z}}{}$&$
\stackrel {\6{\6{\6{\6{\6{\6{\Z}}}}}}\,\6{\6{\6{\Z}}\,\6{\Z}}}{}$&$
\stackrel {\6{\6{\6{\6{\6{\6{\Z}}}}}}\,\6{\6{\6{\Z}}\,\6{\Z}}\,\6{\Z}}{}$&$
\stackrel{\6{\6{\6{\6{\6{\6{\Z}}}}}}\,\6{\6{\6{\Z}}\,\6{\Z}}\,\6{\6{\Z}}}{}$&$
\stackrel{\6{\6{\6{\6{\6{\6{\Z}}}}}}\,\6{\6{\6{\Z}}\,\6{\Z}}\,\6{\6{\Z}}\,\6{\Z}}{}$&$
 \dots
 $
 \cr
  \vrule height 12pt depth 0pt width 0pt&$\!\!^{\Exp 6}$&$\dots$&&&&&&&&&&&\cr
\hline
  &$-$&$+$&$+$&$-$&$+$&$-$&$-$&+&+&$-$&$-$&$+$&$\dots$\cr
$^{\mbox 5}$&
$
\stackrel {\6{\6{\6{\6{\6{\Z}}}}}}{}$&$
\stackrel{\6{\6{\6{\6{\6{\Z}}}}}\,\6{\Z}}{}$&$
\stackrel{\6{\6{\6{\6{\6{\Z}}}}}\,\6{\6{\Z}}}{}$&$
\stackrel{\6{\6{\6{\6{\6{\Z}}}}}\,\6{\6{\Z}}\,\6{\Z}}{}$&$
\stackrel{\6{\6{\6{\6{\6{\Z}}}}}\,\6{\6{\6{\Z}}}}{}$&$
\stackrel{\6{\6{\6{\6{\6{\Z}}}}}\,\6{\6{\6{\Z}}}\,\6{\Z}}{}$&$
\stackrel{\6{\6{\6{\6{\6{\Z}}}}}\,\6{\6{\6{\Z}}}\,\6{\6{\Z}}}{}$&$
\stackrel {\6{\6{\6{\6{\6{\Z}}}}}\,\6{\6{\6{\Z}}}\,\6{\6{\Z}}\,\6{\Z}}{}$&$
\stackrel {\6{\6{\6{\6{\6{\Z}}}}}\,\6{\6{\6{\Z}}\,\6{\Z}}}{}$&$
\stackrel{\6{\6{\6{\6{\6{\Z}}}}}\,\6{\6{\6{\Z}}\,\6{\Z}}\,\6{\Z}}{}$&$
\stackrel{\6{\6{\6{\6{\6{\Z}}}}}\,\6{\6{\6{\Z}}\,\6{\Z}}\,\6{\6{\Z}}}{}$&$
\stackrel{\6{\6{\6{\6{\6{\Z}}}}}\,\6{\6{\6{\Z}}\,\6{\Z}}\,\6{\6{\Z}}\,\6{\Z}}{}$&$
\dots
 $\cr
& $\!\!^{\Exp{5}}$&$\dots$&&&&&&&&&&&\cr
\hline
  &$-$&$+$&$+$&$-$&$+$&$-$&$-$&+&+&$-$&$-$&$+$&$\dots$\cr
$^{\mbox 4}$& 
$
\stackrel {\6{\6{\6{\6{\Z}}}}}{}$&$   
\stackrel{\6{\6{\6{\6{\Z}}}}\,{\6{\Z}}}{}$&$
\stackrel{\6{\6{\6{\6{\Z}}}}\,{\6{\6{\Z}}}}{}$&$
\stackrel{\6{\6{\6{\6{\Z}}}}\,{\6{\6{\Z}}\,\6{\Z}}}{}$&$
\stackrel{\6{\6{\6{\6{\Z}}}}\,{\6{\6{\6{\Z}}}}}{}$&$
\stackrel{\6{\6{\6{\6{\Z}}}}\,{\6{\6{\6{\Z}}}}\,\6{\Z}}{}$&$
\stackrel{\6{\6{\6{\6{\Z}}}}\,{\6{\6{\6{\Z}}}\,\6{\6{\Z}}}}{}$&$
\stackrel {\,\6{\6{\6{\6{\Z}}}}\,{\6{\6{\6{\Z}}}\,\6{\6{\Z}}}\,\6{\Z}}{}$&$
\stackrel {\6{\6{\6{\6{\Z}}}}\,{\6{\6{\6{\Z}}\,\6{\Z}}}}{}$&$
\stackrel {\6{\6{\6{\6{\Z}}}}\,{\6{\6{\6{\Z}}\,\6{\Z}}}\,\6{\Z}}{}$&$
\stackrel {\6{\6{\6{\6{\Z}}}}\,{\6{\6{\6{\Z}}\,\6{\Z}}}\,\6{\6{\Z}}}{}$&$
\stackrel {\6{\6{\6{\6{\Z}}}}\,{\6{\6{\6{\Z}}\,\6{\Z}}}\,\6{\6{\Z}}\,\6{\Z}}{}$&
$\dots$\cr
 &16 &17&18&19&20&21&22&23&24&25&26&27&\dots\cr
\hline 
 &$-$&+&+&$-$&$-$&$+$&$-$&$-$&+&$-$&$-$&+&\cr
$^{\mbox 3}$ & 
$
\stackrel {\6{\6{\6{\Z}}}}{}$&$  
\stackrel{{\6{\6{\6{\Z}}}\,\6{\Z}}}{}$&$ 
\stackrel{\6{\6{\6{\Z}}}\,\6{\6{\Z}}}{}$&$
\stackrel{{ \6{\6{\6{\Z}}} \, \6{\6{\Z}} \, \6{\Z}}}{}$&$
\stackrel{{\6{\6{\6{\Z}}\,\6{\Z}}}}{}$&$
\stackrel{{\6{\6{\6{\Z}}\,\6{\Z}}}\,{\6{\Z}}}{}$&$ 
\stackrel{{\6{\6{\6{\Z}}\,\6{\Z}}}\,{\6{\6{\Z}}}}{}$&$
\stackrel{{\6{\6{\6{\Z}}\,\6{\Z}}}\,{\6{\6{\Z}}}\,\6{\Z}}{}$&$
\stackrel{{\6{\6{\6{\Z}}\,\6{\Z}}}\,{\6{\6{\6{\Z}}}}}{}$&$
\stackrel{{\6{\6{\6{\Z}}\,\6{\Z}}}\,{\6{\6{\6{\Z}}}}\,\6{\Z}}{}$&$
\stackrel {{\6{\6{\6{\Z}}\,\6{\Z}}}\,{\6{\6{\6{\Z}}}}\,\6{\6{\Z}}}{}$&$ 
\stackrel{{\6{\6{\6{\Z}}\,\6{\Z}}}\,{\6{\6{\6{\Z}}}}\,\6{\6{\Z}}\,\6{\Z}}{}$&\cr
 &4 &5&6&7&8&9&10&11&12&13&14&15&\cr
\hline
 &$-$&+&&&&&&&&&&&\cr
$^{\mbox 2}$  &  
$
\stackrel {\6{\6{\Z}}}{}$&$
\stackrel{{\6{\6{\Z}}}\,\6{\Z}}{}$&&&&&&&&&&&\cr
&2&3&&&&&&&&&&&\cr
 \hline
 &$-$&&&&&&&&&&&&\cr
$^{\mbox 1}$ & 
$\stackrel {\6{\Z}}{}$&${}$&&&&&&&&&&&\cr
&1 &&&&&&&&&&&&\cr
\hline
&+&&&&&&&&&&&&\cr
${0}$  &\kern3pt${\Z}$ 
 &&&&&&&&&&&&\cr
&0 &&&&&&&&&&&&\cr
\hline
\hline
&$X$&&&&&&&&&&&&\cr
$r$  &${e_q}$&&&&&&&&&&&&\cr
&$q$&&&&&&&&&&&&\cr
\hline
\end{tabular}
\]
}
\ETA
In Table \ref{T:POLYADICS}, the rank of a polyadic is its height
in bars.
The degree is the number of stacks, as seen from above.
The statistics is $+, -$ for even and odd.

Linear operators $\Deg$, $\Rank: \3Q\to \3Q$
 multiply each element of  ${}^r\3Q^g$ by
$g$ or $r$ respectively.
Let us define the {\em hyperexponential function} $\Exp r$ 
by setting $\Exp 0:=1$ and $\Exp (r+1) :=2^{\Exp r}$.
The dimension of stage ${}_r\3Q=\Grass ^r \1R$ is $\Exp r$.

Table \ref{T:MONADICS} gives the 16 basic monadics $\2e_q$ of rank 4,
 where  $q$ is the serial number of a basis element and $L=\log_2 q$\/.
\BTA [ht]
\caption{Monadics of rank 4}
\label{T:MONADICS}
\[
\begin{array}{c|cccccccccccccccc}
\log_2 q&0 &1&2&3&4&5&6&7&8&{9}&{{10}}&{{11}}&{{12}}&{{13}}&{{14}}&{{15}}\\
\hline
\2e_q&\rule{0pt}{15pt} {\6 {\Y}}& {\6{\6 {\Y}} }&{ \6{\6{\6 {\Y}}} }&{ \6{\6{\6 {\Y}}\,{\6 {\Y}}} }&{
 \6{\6{\6{\6 {\Y}}}} }&{ \6{\6{\6{\6 {\Y}}}\,{\6 {\Y}}} }&{  \6{\6{\6{\6 {\Y}}}\,\6{{\6 {\Y}}}} }&{   \6{\6{\6{\6 {\Y}}}\,\6{{\6 {\Y}}}\,\6 {\Y}} }&{ 
 \6{\6{\6{\6 {\Y}}\,\6 {\Y}}} }&{  \6{\6{\6{\6 {\Y}}\,\6 {\Y}}\,\6 {\Y}} }&{   \6{\6{\6{\6 {\Y}}\,\6 {\Y}}\,\6{\6 {\Y}}}  }&{ \6{\6{\6{\6 {\Y}}\,\6 {\Y}}\,\6{\6 {\Y}}\, \6 {\Y}} }&{
 \6{\6{\6{\6 {\Y}}\,\6 {\Y}}\,\6{\6{\6 {\Y}}}}  }&{  \6{\6{\6{\6 {\Y}}\,\6 {\Y}}\,\6{\6{\6 {\Y}}}\,\6 {\Y}} }&{  \6{\6{\6{\6 {\Y}}\,\6 {\Y}}\,\6{\6{\6 {\Y}}}\,\6{\6 {\Y}}} }&{  \6{\6{\6{\6 {\Y}}\,\6 {\Y}}\,\6{\6{\6 {\Y}}}\,\6{\6 {\Y}}\,\6 {\Y}}}\\
 \end{array}
\]
\ETA
\subsection {\em Interpretation} 
Rays in $\3Q$ and its dual space  $\Dual \3Q$  represent quantum {\sc io} (input-output)   channels
 for the system under study.

$1$  in $\3Q$  represents the io channel for an empty plexus.
$0\in \3Q$ represents the empty channel.

Mathematical objects and classical plexuses, by definition, do not
change under perception. 
Quantum ones on the contrary  change unpredictably, 
in quantities not being perceived.
Thus the quantum simplex or plexus is not a mathematical
object as a classical one might be. 
Its channels, however, due to their macroscopic statistical aspect, can be 
represented by mathematical objects, the elements of
the representation space  $\3Q$ or its dual.

$\3Q$  has no significant relativity group;
 every ray in $\3Q$ is intrinsically different from every other.
This holds also
for $\1R^4$, the representation space for space-time, and
for  the Hilbert space $S^2(\1N)$
 of complex sequences of summable square, the representation space
 for canonical quantum theory.
  Let us take this categorical describability as a prerequisite for a representation space.
 The relativity groups of the physical system preserve selected
 structural elements of the representation space and violate others.
Here  as in canonical theories the linear algebraic operations 
 $+, \vee$  on the representation space are supposed to have physical meaning, and
 the relativity group of the plexus respects them, but violates $\iota$\/.
 There is such an $\iota$ violation in Hilbert's $S^2(\1N)$ too:

 Whatever the statistics of $Q$,
$\iota Q$ has odd statistics.
 This  curious violation of statistics seems acceptable, since we see conservation of statistics only on the particle rank.
It is already implanted in quantum field physics.
All the components of a spinor, 
regardless of their degree as Grassmann products of semivectors,
are given a fresh degree of 1;
and  the Dirac vector $\gamma^{\mu}$, of even spin parity,  obeys odd commutation relations with itself.
Nevertheless $\iota$  preserves the spin-statistics correlation, by
violating spin parity exactly as much as exchange parity.

\section{The spinor tree}
 $\3Q$ is  constructed on the basis
of Fermi statistics, spinor theory, and classical set theory.
The spin-statistics equality $W = X$ is a major clue to its interpretation.

Since monadics in $\3Q^1$ anti-commute, 
vertices obey the exclusion principle, and have $X = 1$.
Therefore, since $W=X$,  monadics must be spinors, with $W = 1$. 
Vertices must have spin 1/2, and $\3Q$ must consist of multispinors of various 
degrees and ranks and their superpositions.
The theory of the  algebra $\3Q$ must then include
the quantum theory of Fermi statistics, and the classical theory of spinors.

The theory of spinors is commonly based, following Cartan and Dirac,
on  a classical space-time background.
Presumably there is no classical space-time.
Then we need another interpretation  for spinor theory.
 The spin-statistics equality $W = X$ suggests one.
When Cartan (1913) constructed double-valued representations for rotations, 
 Schur (1911) had already constructed them for permutations, 
  a more primitive construct, using much the same matrices, among others.
   
The spin-statistics correlation can then be made into an identity.
The classical angular-momentum Poisson Bracket relations are a singular classical limit of  a regular quantum statistics of the Palev (1977)  kind, reviewed in \S \ref{S:PALEV},
which is taken as more fundamental.
This statistics in turn follow from Fermi statistics for quanta  that are
exchanged by spin operations.

\subsection  {\em The quadratic space $\3W$} 
\subsection {\em Indefinite metric}

Cartan constructed his spinor space as the exterior algebra over a ``semivector" space.
This step in the construction uses no metric.
Let us identify it with the construction of each 
stage of $\3Q$ from the preceding.
Thus each stage serves as  semivectors for the next and spinors for the preceding;
the constructs of semivector and spinor are relativized.
{\em The spinor space of any rank is the exterior algebra over that of the previous rank.}
 $\3Q$ can be interpreted either as a spinor space, a modularization of 
the infinite-dimensional spinor space of Dirac (1974), 
or as a semivector space.

In the Cartan theory, a quadratic vector space is prior to the spinor space and defines its orthogonal group.
It  is the duplex space of the Cartan
 semivector space.
 Here therefore the quadratic space of stage $r$  is 
 \BEQ
 {}_r\3W:=\Dup {}_r\3Q\/, \quad \|v\oplus v'\|_{\4D}:= v'(v)
 =v'\circ v\/.
 \EEQ
 {\em The quadratic space $ {}_r\3W$ for  spinors of any stage  $r+1$ is the duplex space 
 of the spinors of the previous stage.}
Its orthogonal group $\SO(\Exp r, \Exp r)$  includes the linear group $\SL(\Exp r)$
of the spinors of the previous stage.
The duplex space is a representation space for a pair of a simplex and a dual simplex, which will be called a {\em duplex} for short.
It carries a neutral quadratic form that can serve as the origin of the Minkowski metric
of space-time tangent spaces. 
It is encoded in Dirac operators $\gamma_w$ for the duplex.
This Clifford algebra is isomorphic for all duplexes of a given rank, 
as for all tangent spaces of Minkowski space-time.
It varies at the event level because the cell assembly is  a variable.
 
The spinor space is also duplicated in practice.  If $\3Q$ is is used for kets then $\Dual \3Q$ is used for bras, and $\Dup \3Q$ contains superpositions of bras and kets.
These do not arise in canonical quantum theories.
This can be regarded as  a superselection rule resulting from a singular limit 
of large numbers at the level of the observer.

The statistical norm $\|Q\|$ for any vector $Q$ of the representation space
can be normalized to be the average number of systems flowing into
the experiment in the channel represented by $Q$. 
Then the norm $\|Q\|$ is positive for kets, negative for bras;
it is no longer a probability but a probability flux.
Perhaps Dirac (1974) intended this interpretation of indefinite  metrics.\nocite{DIRAC1974}
Transition probabilities are homogeneous of degree 0
in the norm, so that  no
physical conclusion is changed if 
a norm is replaced by its negative everywhere.
Processes associated with bras are defined to lower the energy; kets raise the energy.
The energy distinction does not depend on the sign of the norm
and therefore can be used to define it.

Null vectors in $\Dup \3Q$ are
 inaccessible in the singular canonical limit, 
like null vectors of space-time in the Galilean limit.

\subsection{\em Dirac spin operators} 
 The quadratic space and the spinor space are linked by an algebra ${}_r\3A$, 
which is both the Clifford algebra 
 of the quadratic space and the operator algebra of the spinor space.
 Here therefore the Clifford algebra of stage $r$
 is
 \BEQ
 {}_r\3A:=\Cliff {}_r\3W= \Cliff \Dup \iota\apost {}_r\3Q\/.
 \EEQ

{\em The operator algebra of rank $r$ is a Clifford algebra over the duplex space of the previous rank, taken with the duplex metric}\/:
\BEQ
 \Alg \22^{\3V} = \22^{\3V} \ox \22^{\Dual\3V}=\22^{\Dup \3V} = \Cliff \Dup\3V\/.	
  \EEQ
  Since the elements of this Clifford algebra are the linear operators on $\Grass {\3V}$\/, $\Grass{\3V}$
   is
indeed a spinor space for this Clifford algebra and its orthogonal group.

In particular,  each basis vector $\2e_w \in {}_{ r-1}\3W$
may be identified  with a Dirac operator,
a generator  of a Clifford
algebra of the next rank, when it is usually written as $\gamma_w$.
Dropping the  rank prescript for now,
we 
define the operator $\gamma_v$ to be the  left exterior multiplication by $\2e_v$, 
 and $\gamma^u$  to be the left exterior differentiator $\partial^u$ with respect to $\2e_u$. 
These  indeed generate $\Cliff ({}_r\3W)$:
\BEQ
\{\gamma_u, \gamma^v\} =\delta^v_u,\quad  \{\gamma^{v'}, \gamma^v\} = 0= \{\gamma_u', \gamma_u\}\/.
\EEQ
The usual spin operators generating the spin group $\Spin(\3W)$ are the semi-commutators
\BEQ
\gamma_{w'w} := \frac12 [\gamma_{w'},\gamma_w]\/.
\EEQ
A rotation  of a simplex through $2\pi$  is represented by 
\BEQ
e^{iW}=e^{i\pi\Sigma\gamma_{w'w}}
\EEQ
where the sum is over the vertices.
The {\em spinor tree}\/, the central column of  Table \ref{T:TREE},  is constructed inductively, 
by iterating $\Grass $,
 beginning with the trivial exterior algebra $\1R$ of degree 0 and rank   0. 
 Orthogonal groups and Clifford-Fermi algebras perch on each level of the resulting  spinor tree of Table \ref{T:TREE}.
 \BTA
\BEQ
\label{E:TREE}
\begin{array}{|c|c||c||c|c|c|}
\hline
\vdots&\vdots&\vdots&\vdots&\vdots\\
&&\uar\mbox{\kern-4pt\sm ext}&&\\
3&\Fermi 4 \1R&16 \1R&32\1R&\SO(16,16)\\
&&\uar\mbox{\kern-4pt\sm ext}&&\\
2&\Fermi 2\1R&4\1R&8\1R&\SO(4,4)\\
&&\uar\mbox{\kern-4pt\sm ext}&&\\
1&\Fermi \1R&2\1R&4\1R&\SO(2,2)\\
&&\uar\mbox{\kern-4pt\sm ext}&&\\
0&\Fermi 0&\1R&0&1\\
\hline
r & \Alg {}_{r }\!{\Q} & {}_{r }\!{\Q}&_r \3W{} & \SO(_r\3W )\\
\mbox{\sm Level}& \mbox{\sm  Algebra} & \mbox{\bf \sm Spinors} & \mbox{\sm Vectors}&\mbox{\sm Group}\\
\hline
\end{array}
\EEQ
\caption{\bf  Stages of the spinor tree $\3Q$. }
\label{T:TREE}
\ETA
The symmetry Lie algebra of ${}_{r }\!{\Q}$ as  vector space is $\slin({}_{r }\!{\Q})$\/.
The symmetry Lie algebra of ${}_{r }\!{\Q}$ as exterior algebra is only $\slin({}_{r-1 }\!{\Q})$,
exponentially smaller.

\subsection{\em  Pauli metrics}
 When monadics in $_r\3Q$  serve as  spinors, 
let us endow them with a recursive  {\em Pauli metric} form $_r\beta$
  that is invariant under the spin group $\Spin(_r\3W) $.
 The usual Pauli form is stage $r = 2$.
Dropping the prescript $r$ for clarity,
$\beta$ is constructed to define a pseudo-expectation value
\BEQ 
\Av_Q  A :=  Q^{ {}\beta} AQ ={}\beta Q A Q ={}\beta_{q'q''} Q^{q'}A^{q''}{}_q Q^q
\EEQ
giving the flux of any variable $A\in \Alg {}_r\3Q$ for any $Q\in {}_r\3Q$.
This transforms in the same way as  $A$ under the spin group of stage $r$.

This requires that the Pauli metric $\beta$ of each Clifford algebra $\Cliff \3W$ 
skew-symmetrize all the $\Gamma=\gamma_{w'w}\in\Cliff {}_r\3W$,
 in the sense that 
(designating the transpose operation by $\4T$)
\BEQ
\forall \Gamma \in \Cliff^2\3W :  \4T(\beta \Gamma ) = -\beta\Gamma\/.
\EEQ

 The construction of $\beta$ for $\SO(n,n)$
 then follows the usual one for $\SO(3,1)$ closely.
 To skew-symmetrize the second grade it suffices to skew-symmetrize the first grade. 
 Call a basis for $\3W$ {\em orthonormal } if the duplex metric $\4D$ is diagonal in that basis
and has diagonal elements $\pm 1$.
 One matrix representing a Pauli metric $\beta$ in an orthonormal basis 
 is the product (in any order) of those $\gamma_w$ whose squares are $-1$.
Metrics and operators transform differently; hence the frame restriction.

The standard Dirac case   $_2\beta$ is skew-symmetric, but $_r\beta$ 
 is symmetric for $r > 2$.
The Pauli metrics ${}_r\beta$do not have a well-defined limit as $r\to \infty$.
Since the stages  nest, however, so do their groups, and the $\beta$ 
of any stage serves as well for all lower stages.

\section{Chiral spinors}
\label{S:CHIRAL}

\subsection{\em Infinities} 
 In classical mechanics the center of the coordinate Lie algebra
 is the whole algebra.
 Canonical quantization reduces the center to the one-dimensional
 Lie algebra $\1C$  of complex numbers, generated by the right-hand sides of the canonical commutation
 relation
 \BEQ
 \label{E:CCR}
  [ip,  iq] = i; \quad [i,  ip] = 0; \quad [iq, i] = 0; 
  \EEQ
 in which $ip, iq,  i $  are skew-hermitian infinitesimal isometries of the quantum theory.
Representing the $ i $ in (\ref{E:CCR}) by a scalar matrix is well known to make the Lie algebra singular;
   in a regular theory in a complex  $n$-dimensional representation space the trace of the first equation would be $ 0 =n i$\/. 
   
   These canonical commutation relations, in their many appearances, 
   from the differential calculus to the  Bose statistics of gaugeons,
   seem to be the only obstacle
   to regularity in all of quantum field theory.
 They account for  all the infinities of present-day
   quantum physics.
   Therefore their reform is a critical precondition for a regular theory.
Yang (1947) reformed the canonical commutation relations between $x^{\mu}$ and $p_{\mu'}$,
and Palev (1977) those between bosonic creators and annihilators.

The Yang (1947)  Lie algebra of (\ref{E:QSPACETIMES}) 
replaces
  the  $i$ in  (\ref{E:CCR}) by a generator  $L^{65}=:L_i$ of    the Yang Lie algebra,  here taken with the neutral signature of $\so(3,3)\subset \slin(4\1C)\subset \slin(8\1R)$.
Its spinor representation space is a subspace of cellular stage ${}_C\3Q$, with rank $C$.
  It seems that an 8-dimensional spinor space with $C=4$ suffices for this.
  This cell will be adequate for the gauge theories of gravity and electricity only.
 The Standard Model requires a cell of perhaps 16 vertices,
 which will still fit in stage 4.
  
\subsection{\em Higgs field and $i$}
Suitably renormalized, one generator $L_i$ becomes a quantized imaginary $\Qi$, 
which becomes $i$
in a suitable singular limit with organization.
Let us therefore adopt the real field $\1R$ for the coefficients of the representation space $\3Q$.
 This makes it necessary to replace the Schr\"odinger first-order differential equation
 for the time-development of input vectors, with its prominent $i$, by a real one,
that does not require an
 explicit $i$, like Dirac's  or Maxwell's equation;
 though it is understood that the canonical action principles for both 
 Dirac's  and Maxwell's equations presently include  essential $i$'s that must now be
 replaced by an operator like $\Qi$.
 In various limits, stages, and normalizations, Yang's $L_i$ is supposed to underlie the $i$ of quantum probability amplitudes, the 
 electric charge gauge generator $iq$, and the Higgs field.

{\em The dynamical symmetry rearrangement  that centralizes the imaginary $i$ 
reduces general spinors of the Yang group to chiral spinors of the Lorentz group.}

We see this as follows.
The chirality  of a fermion in the standard quantum theory
is an operator at the top of its  16-dimensional Dirac Clifford algebra,
\BEQ
i\gamma^{4321}:=i \gamma^4\gamma^3\gamma^2\gamma^1\doteq \pm 1=:i \gamma^{\top}\doteq \pm 1\/.
\EEQ
$i\gamma^{4321}\doteq 1$ for a left-handed electron, with isospin 1/2,
 and $-1$ for a right-handed electron, with isospin 0.

The Yang $\SO(3,3)$ group  has a spinor space $\cong  8\1R$, the square root of its Clifford algebra.
Its pseudoscalar volume element $\gamma^{\top}:= \gamma^{654321}=\gamma^{65}\gamma^{4321}$ 
commutes with $\so(3,3)$ transformations $\gamma^{y'y}$ 
($y,y'=1,\dots, 6$)
and
has eigenvalues $\gamma^{\top}\doteq \pm 1$.  
$\gamma^{\top}$  reduces the spinor space to two eigenspaces $\cong 4\1R$ with 
\BEQ
\gamma^{4321}=\mp \gamma^{65} =\mp \gamma\Qi\/.
\EEQ
These are therefore chiral spinors.

The atoms of the quantized orbital operators
may then be represented by 
$\delta \8x^m\sim \gamma^m$, $\delta \8p_m=\gamma^{4321}\gamma_m$,
as was suggested also by Marks (2008) \nocite{MARKS2008}.

The canonical relations work
in a part of the  spectrum of 
$|\Qi|=+\sqrt{-\Qi^2}$ that is so near to the maximum value $ \4N$
as to be indistinguishable from it,
yet has passed for infinite until now.
For example the band
\BEQ
1 - {\4N}^{-1/2}<  |\Qi| \le  1\/
\EEQ
is sufficiently narrow and crowded,
with width ${\4N}^{-1/2}\to 0$ and
 multiplicity $\0O( \sqrt N)\to \infty$.
 
 St\"uckelberg (1960) showed 
 how to extract the complex quantum theory from a real one
like $\3Q$  given an operator $\Qi$ whose square is $-1$.\nocite{STUECKELBERG1960}
 The main point is that 
 when the superselection rule for $\Qi$ is in force, 
physical io channels are not represented by 
 vectors in $\3Q$ but by planes in $\3Q$ invariant under $\Qi$.
For physical operators must then commute with $\Qi$ and 
no one-dimensional projector can do this, while a two-dimensional one can.
 Two vectors  $Q_1, Q_2\in \3Q$ with $Q_2=\Qi\, Q_1$ define such a plane,  
 and  combine into one ray in a singular limit where $\Qi\cto i$ is adjoined to the real field.
 The origin of this superselection rule was unspecified.
  In the case of a plexus, it arises from the quantum law of large numbers,
 as for all the coordinates of classical mechanics.
 The imaginary $i$ is the singular limit $\Qi$, the sum of $\4N$ cell operators $\delta\Qi$,
 all much smaller than unity, and commuting with one another.
 The other basic variables are similar sums.
 The commutator of a variable with $i$ then has only $n$ terms, while the product has 
 $\sim n^2$.  
 For large enough $n$ the commutator is negligible compared to the product.

\subsection{\em Cumulation}  
\label{S:CUMULATION}

A vertex monadic of any stage
is a superposition of unitized products of  monadics of the previous stage.
Thus low-level operators 
induce high-level ones.
The {\em  cumulation operator} $\Sigma$ converts any property $x$ of a module
of low rank into a property $\Sigma x$ of a module one rank higher,
the {\em cumulant} of $x$,
by summing over all replicas of the lower-rank module in the higher.
Kostant (1961) calls it   $\theta$. \nocite{KOSTANT1961}
In particular, in  a quantum plexus
the quantum spin operators of a single-cell stage  induce orbital operators
on the event stage by 
  $\Sigma$,
often  written in terms of  annihilation and creation 
operators $ \psi$ and $ \psi^{\4H}$ as 
\BEQ
\Sigma x:=\psi^{\4H} x \psi\/.
\EEQ
$\Sigma$ is a Lie homomorphism.
It injects the Lie algebra of each rank
into that of the next.
$\Sigma^n x$  represents any operator $x\in \Alg({}_{r}\!{\3Q})$ in the algebra $\Alg({}_{r+1}\3Q)$.
Let us also
write $x=\delta y$ to mean that $y=\Sigma x$, and $x=\delta^n y$ to mean that $y=\Sigma^n x$.

\subsection{\em Single-cell symmetries} Not all of the generators of Yang $\so(6-n,n)$ are symmetries of the lamina, evidently.
Our ambient space-time lamina is not six-dimen\-sional on our macroscopic scale.
But its simplicial cells might be six-dimensional on their sub-microscopic scale.
We can use the Yang group generators  as symmetries of the single-cell stage, 
when we write them as $\delta L_{n'n}$.
Their macroscopic cumulants are $L_{n'n}:=\Sigma^{E-C} \delta L_{n'n}$.
Then
\BEQ
\label{E:YANGALG}
[\delta L_{n'n},\delta L_{m'm}] =g_{nm'}\delta L_{n'm}- g_{n'm'}\delta L_{nm}+g_{n'm}\delta L_{nm'} - g_{nm}\delta L_{n'm'}\/,
\EEQ
with $m,m',n,n'\in \{1,\dots, 6\}$.
The six-dimensional quadratic Yang space requires a three-dimensional semi-vector space  $3\1R\subset {}_C\!\3Q$ of the cellular rank,
and its eight-dimensional spinors are elements of the eight-dimensional $\Grass  3\1R$.

Relative to any frame, these carry atomistic elements of familiar variables according to
\BEQ
\label{E:YANGSPACE}
\delta L=
\left[
\begin{array}{ccc|cc}
 0&\dots&\delta  L^{14}&-\delta x^1/\4X&-\delta p^1/\4E\\
  \delta L^{21}&\dots&\delta  L^{24}&\vdots&\vdots\\
  \vdots & & \vdots & \vdots&\vdots\\
\delta  L^{41}&\dots&0&-\delta x^4/\4X&-\delta p^4/\4E\\
\hline
\delta x^1/4X & \dots&\delta x^4/\4X &0 &\delta L^{56}\\
\delta p^1/\4E & \dots &\delta p^4/\4E &\delta L^{65}&0\\
\end{array}
\right]\/,
\EEQ
whose matrix elements are elements of $\so(6-n,n)$.
Here $\4X$ and  $\4E$ are fundamental units of time and energy,
the {\em chrone} and the {\em erge};
it is not excluded that these have the order of magnitude of Planck units.
$\delta L_i=\delta L^{56}$ is one atom in the cumulant 
 $\Sigma \delta L^{56}=\4N\Qi$,
 $\Qi$ is a quantized $i$, and $\4N$ is the number of terms in the sum.

On the other hand, the commutation relations  
of the Heisenberg-Poincar\'e Lie algebra of space-time position $x^{\mu}$\/,
momentum-energy $p^{\mu}$\/,  
and dimensionless Lorentz generator $L_{\mu'\mu}$\/,
in $c=1$ units,
are
\BEA
\label{E:HPA}
[L^{\nu'\nu},L^{\mu'\mu}]&=&g^{\nu'\mu'}L^{\nu\mu}-  g^{\nu\mu'}L^{\nu'\mu}+ g^{\nu\mu}L^{\nu'\mu'}-g^{\nu'\mu}L^{\nu\mu'}\/,\cr
[L^{\nu\mu'}, x^{\mu}]&=&g^{\mu'\mu}x^{\nu}-g^{\nu\mu}x^{\mu'}\/,\cr
[L^{\nu\mu'}, p^{\mu}]&=&g^{\mu'\mu}p^{\nu}-g^{\nu\mu}p^{\mu'}\/,\cr
[x^{\mu'},p^{\mu}]&=&i\hbar g^{\mu'\mu},\cr
[L^{\nu\mu'}, i]&=&
[x^{\mu'},x^{\mu}]\;=\;
[x^{\mu},i]\;=\;0
\EEA
with $\mu,\mu',\nu,\nu'\in \{1,\dots,4\}$\/.
To contract (\ref{E:YANGALG}) $ \cto $ (\ref{E:HPA}),
we must polarize and centralize one of its generators, 
which we may take to be
$L_{56}$ in an adapted frame,
and turn off the gauge fields.
Let us invoke a dynamical self-organization akin to magnetization.
It is sufficient if (\ref{E:YANGALG})  acts
on an io space $\3V=2(2\4N+1)\1R\subset \3Q$ with large
dimension  $2(2\4N+1)\gg 1$, 
so that the spectra of  the $L_{m'm}$ 
are quasi-continuous;
and  if in a polarized sector 
$\3V_{\rm pol}\subset \3V$, 
supposed to represent the usual environment,
$|L_{65}|$ is close to its maximum eigenvalue $\4N$,
and the other components of $L_{\mu'\mu}$
are much smaller, though still quasicontinuous:
\BEQ
|L_{65}|\approx \4N \gg |L_{\mu'\mu}|\/.
\EEQ

The matrices of the single-cell $\so(\3W[C])$ are not  orbital variables but
their ``atoms'',  spin variables.
The orbital variables are their cumulants on a later stage.
A Yang $\so(3,3)$ of level $C$  is faithfully represented on level $E$
by second cumulants of its generators $L^C$. 
The $L^C$ in one $\Q$ frame, up to constant multipliers, approach the operators
\BEA
\label{E:QYANG}
\begin{array}{lll}
\delta \8 x^m &=& \4X\,\gamma^{m5},\quad m,n=1,2,3,4 \/,\cr
\delta\8 p_m& =& \4E \,\,\gamma_{m6}\/,\cr
\delta\,\Qi &= &\gamma^{65}\/ , \cr
\delta\8L_{nm}&=&h \gamma_{nm}\/, 
\end{array}
\EEA
as $ \0h(4) \cfro    \so(3,3)\cfro \slin(6)$\/.
The quantized cumulant imaginary $\Qi$ is normalized to unit magnitude with a small  factor $\4N^{-1}$.
To form macroscopic event coordinates, 
we  must cumulate these single-cell variables at least twice,
to
reach at least
level 6, with
 $\Exp 6=2^{(2^{16})}$ points, enough for a quasi-continuum.
Let us set
\BEA
x^{\mu}  &=&  \4X L^{\mu 5}\/,\cr
 p^{\mu} &=&   \4N^{-1}\4E L^{\mu 6} \/,\cr
\Qi   &=&  \4N^{-1} L^{65},
\EEA
following Yang.
$\Qi^2\approx -1$ in $V_{\rm pol}$\/.
The factor $\4N^{-1}$ results from the duality between position and momentum,
which requires us to average one when we sum the other.

In the limit
\BEQ
\label{E:YANGLIMIT}
\4N\to \infty,\quad \mbox{ with } \4E\4X=\hbar,
\EEQ
the Heisenberg-Poincar\'e relations (\ref{E:HPA}) follow.

The electric generator $\delta L_{56}$ commutes with the Lorentz generators $\delta L_{\mu'\mu}$
(in an adapted frame) at the cellular level
but not with the atoms of
momentum $\4E \delta L_{\mu 6}$ or position $\4X \delta L_{\mu 5}$\/.
That commutativity comes at the event level in the lamina, due to the large value of $\4N$.
%

\subsection{\em Palev statistics}
\label{S:PALEV}
The commutation relations for Bose statistics define a canonical Lie algebra,
and so they too require regularization.
 Palev (1977), pursuing simplicity,
reformed the canonical algebra of 
Bose statistics to  a simple Lie algebra, such as $\so(N)$.\nocite{PALEV1977}

Pairs of $\3Q$ fermions do not exactly obey Bose statistics but
 a Palev statistics
defining an $\so(N_+,N_-)$ Lie algebra.
The canonical Lie algebra of Bose statistics is a singular organized limit
of this $\so(N_+,N_-)$ with $N_{\pm}\to \infty$.
In a recursive quantum  theory  it is natural to regard
all empirical bosons as approximations to palevons
with even-degree 
fermionic cores.
In a canonical quantum theory a fermionic core would show up as a hard core in high-energy collisions.
In the recursive quantum theory this does not follow,
because  at the cellular level the canonical indeterminacy relations have not yet set in, 
and because elementary energy transfers are local in both position space and momentum space.

\subsection{\em Vertex statistics}
\label{S:QSPACES}
To be sure, these event spaces are too small for the lamina  variables of \S \ref{S:LAMINA}\/.
A recursive quantum theory must fit  all these variables 
and their commutation relations
into the operator algebra of $ \3Q$\/.
All these new groups, however, are simple and fit readily into 
low ranks of $\3Q$.
The infinities come from the orbital and bosonic field variables,
which generate singular Lie algebras.
This study limits itself to this core problem.

The first models in (\ref{E:QSPACETIMES})  do not impose odd statistics 
on their space-time atoms or {\em chronons}\/ for obvious reasons.
If the chronon has spin 1/2 and no other variables,
odd statistics would exclude histories with more than four  events.
In the recursive quantum theory, unitization allows us to clone a spin
and build  histories with arbitrarily many spins 1/2 with odd statistics.

\section{ Recursive gauge}

The  quantum plexus  must have a regular gauge group that can be approximated by
the singular ones of gravity and the Standard Model.
Historically, gauge theories were made by gauging un-gauged theories.
Recursive quantum theories, however, are born gauged.
The event stage contains enormous numbers of clones of the single cell. 
The  plexus  gauge group of the first (Dirac) kind  is the group $\SL(n)$ of a single cell, a simplex
some two ranks below the space-time event-vertex.
 The gauge group of the  second kind is that generated by all the single-cell groups of the event rank.
 In the laminar phase, 
  it is to
reduce to that of
the Standard Model and gravity,
first by dynamical symmetry re-arrangement by the organization of the lamina, 
which results in a semi-simple group,
and then by
 a singular approximation, including the limit of classical space-time,  that results in the usual singular gauge group.

Evidently quantization and gauging are both forms of quantification, acting at different stages.
{\em The  single cell  is the gauge block in the Weyl sense.}
It defines a fundamental quantum of time-interval as well as Planck's quantum of action.

This unanticipated unification of the quantum and the gauge fits into a long-standing conjecture:
 that gauge theory, including gravity, is an 
extension of  dislocation theory 
to a quantum crystalline medium;
that every  gauge current seen close up is a Burgers-Volterra vector describing dislocations.
The medium is the quantum plexus, organized into a crystalline polarized lamina, and the gauge fluxes, interpreted as vertex permutations rather than rotations, 
characterize its dislocations.

To see this unification in more detail let us return to classical mechanics. 
There one has  three sets of  algebraic commutation relations, schematically
$[x,x]=0$ , $ [x,p]=0$, and $[p,p]=0$.
This trivial Lie algebra is maximally singular, in that its Killing form is not merely singular
but  vanishes identically.

A gauge theory  reforms the commutators $[p,p]$, making them into a gauge 
curvature field.  
A  canonical quantum theory reforms the commutators $[x,p]$  making them
$i\hbar$.  
Let us define the power  constant $\4W:=\4E /\4N \4X$.
In a recursive quantum theory, the lowest-order corrections to the commutators have the form
\BEQ
[x,x]\sim \frac{\hbar}{\4W}, \quad [x,p]\sim \hbar, \quad [p,p]\sim \hbar \4W.
\EEQ
Thus classical gauge theory handles the $o(\4W)$ correction, canonical quantum theory
deals with the $o(1)$ correction, and the $o(\4W^{-1})$ correction, 
presumably the smallest in ordinary experiments, is not yet taken up.

This symmetry is  masked at first because the $[p,p]$ commutators are field variables while the $[x,p]$ commutators have passed for constants.
In a quantum plexus the $[x,p]$ commutator is also a field, the Higgs field, a singular normalized limit of $L_{i}$.
This passes for constant in ordinary circumstances because it is  polarized near its maximum value by the
dynamical symmetry-rearrangement that forms the lamina.

A gauge connection associates a Lie algebra element $a$ with a direction $p$ (or its dual)
at a point $x$.
A  direction in Yang space (\ref{E:YANGSPACE}) {\em is} 
a Lie algebra element.
This permits us to assume 
that a singular quantum gauge 
field might actually be an organized singular limit  of a regular sea of events,
 taken from a quantum space like a Yang space
with an  appropriate orthogonal group.


$\Q$ degree counts vertices of the  simplex, $\Q$ rank  counts nested iotas,  
and the basic $\Q$ dynamical operators 
$L^{C}\in \so( \3W[C])$ of the cell rank $C$
count components of generalized angular momentum in units of the roots of this Lie algebra.
The eight spin-like atoms of orbital angular momentum  $\8x^m$ and $\8p_m$ ($m\in 6$), 
and the quantized imaginary $\Qi$,
are among the 15 generators $\gamma^{C}$ of a Yang $\so(3,3)$ Lie algebra, forming
a Lie subalgebra of  the 120-dimensional Lie algebra 
$\so(\3W[C] \subset \Alg{}^{4}\!{\Q}{}^1\cong \Cliff \3W{[C]}$ of  rank $C$\/.

The elementary momenta do not commute.
 Neither do the cumulative momenta, which correspond to
  infinitesimal translations.
This quantum non-commutativity survives into general relativity as part of the curvature,
perhaps including a cosmologically constant part.

The Higgs field is usually posited ad hoc.
The quantized $i$ of Yang
 provides it with a theoretical foundation,
 as in Tavel (1965)
 (which flowed from a suggestion of Yang at a Rochester Conference).
 .\nocite{TAVEL1965}
 It seems consistent with the technicolor hypothesis of Susskind (1979) and Weinberg (1979) 
that the Higgs field, like the BCS pair {\sc io} vector,
be an order parameter of a self-organization,
namely that of 
 the crystalline laminar condensation whose disolocations are
 responsible for gauge interactions.
 \nocite{WEINBERG1979, SUSSKIND1979}

 The Yang group and the Heisenberg-Poincar\'e group include orthogonal transformations 
 that interchange space-time position and momentum, a symmetry called reciprocity by Born (1949).\nocite{BORN1949}
Reciprocity is badly broken by 
\BEN
\item The canonical locality principle. This requires elementary particles in interaction to have
approximately the same position, 
but allows them to have widely different momenta;
 especially
by its sharpest form, 
\item The gauge principle.
 This  requires the local gauge group to
 be replicated at every position, not every momentum.
\item The vacuum.
This has no natural zero for position in space-time  
 at the particle level of experiments, but has a natural zero for momentum-energy.
 \EEN
 The recursive gauge principle merely requires covariance under the group of every simplicial cell, which includes 
changes in momentum as well as in position.
 
 Locality in momentum and asymptotic freedom both limit high-momentum transfers
in elementary particle interactions, so they may be related.
One rather tenuous relation arises at once.
Simplicial locality leads in a singular limit to a non-abelian gauge.
When this is adjusted to fit the Standard Model,  it will permit the
Gross-Wilczek deduction of asymptotic freedom,  which seems to be a mild form of locality in momentum, from the non-abelian gauge of the Standard Model.

In a quantum plexus,   the structures treated as elementary particles in the canonical theory
are  edges of extended dislocations
 and their interactions consist of many simplicial processes.
Then the observed large momentum transfers $p^{\mu}$ may occur in many small steps
 $\delta p^{\mu}$ consistent with simplicial locality.
 These violations of reciprocity are trying to tell us something important about the 
 fine structure of the lamina,
 but we have not made out what they are saying yet.
 
\subsection{\em Dynamics}
Let us consider other problems that a recursive quantum electrodynamics 
 must still solve.
 The Yang group suffices for its cellular gauge group,
 with  $L_{56}\cto\4Ni$  as the electric axis in Yang space, 
 the Higgs field to be frozen, and the electric gauge group generator.
The $i$ of the event level is to be treated as a constant. 
 While $\4Ni$ rotates position into momentum, the limit $\4N\to \infty$
 permits us to take $[i, x^{\mu}]=0$\/.
 
 The simplicial operators for the orbital variables have already been given.
 Since the momentum-energy variables $\8p_{\mu}$ are covariant under the simplicial gauge group,
 they correspond to the kinetic momentum-energy $p_{\mu}-\Gamma_{\mu}$
 of the canonical gauge theory rather than to the total momentum-energy $p_{\mu}$
or the potential momentum $\Gamma_{\mu}$, which includes both the electric and the gravitational vector potential.
 No plexus expression  has been found  for either $p_{\mu}$ or $\Gamma_{\mu}$ separately.  
 Perhaps the search was misguided.
 Since only $p_{\mu}-\Gamma_{\mu}$ has invariant physical meaning,
 the absence of separate $p$ and $\Gamma$ 
 makes the recursive theory more physical, not less, than the canonical one.
 To recover pure quantum electrodynamics the
gravitational vector potential  can be suppressed
 by setting 
 $[\gamma^{\mu}, p^{\mu'}]\to 0$\/.
 
 The Lorentz group corresponds to the centralizer of the electric group,
 with generators $L_{\mu'\mu}$ in an adapted frame.
Thus dynamical symmetry rearrangement and a singular limit
have
to account for the  vacuum averages of
(1) $p^{\mu}\cto 0$,  
(2) $\Qi\cto i$, the  imaginary constant and the Higgs field
and (3)  $\6g\cto g_{\mu'\mu}$, the  gravitational field,
as order parameters.
This brings in rather new considerations, appropriate for a sequel.

\section{Discussion}

If we take both the quantum and gravity theories seriously, space-time is
more likely to be composed of spins, as Feynman and Penrose have suggested, than conversely.
 The geometric meaning of spinors is mysterious, as Atiyah (1998) points out.\nocite{ATIYAH1998}
The quantum meaning
 is  simple:  They represent the most elementary quantum actions, the creation and annihilation
of vertex elements, which are of the Fermi-Dirac kind.
Their Clifford algebra is a Fermi-Dirac algebra.
Spinors cannot be explained in simpler terms if there are none.
They  can still be understood by  describe everything else in terms of them, including 
 geometry, as attempted here.
 In physics as in archeology, the foundations are the last thing discovered.
 Classical geometry and mechanics are explained by quantum geometry and mechanics,
 not conversely.

Iterated, the classic spinor construction of Cartan leads to a 
recursive hierarchy of spinors that can be interpreted as 
representations of 
 recursive quantum simplicial complexes with symmetries
that are useful for quantum theories. 
The semivectors underlying the Cartan construction of spinors are also spinors, but
they are of the previous stage, and are given a different transformation law under the orthogonal group, yet to be explained physically.
The neutral metric of
the Cartan construction is now a relativistic probability flux metric 
that distinguishes bras from kets in each
frame by the signs of their norms, 
and reduces to the classical Minkowski metric in an organized singular limit. 

Canonical quantum theories imagine a manifold of events without spins. 
Spinless events do not occur in the Standard Model.
The events of a recursive quantum theory are  vertices of a simplex
and carry the spin of its group.
A complex of such vertices, as opposed to a lattice, is finite in that
its dynamical variables all have discrete finite spectra.

Manifold theories represent the event space, including event coordinates that suffer gauge transformations,
 as a manifold, preferably  without boundaries,
after Kaluza.
Then they need a high energy-density and stress to curve and compactify 
the circular transverse dimensions of the lamina.
Humbler laminas like
bubbles, snowflakes, and graphenes
do not   compactify but organize.
Their quantum elements organized themselves in only some of the possible directions,
typically because their interactions  have short range and saturate, 
like covalent bonds;  and so they have boundaries in the transverse dimension.
Let us assume the same for the ambient laminar plexus.
This replaces the compactification problem by the organizational problem:
How do the available modules assemble themselves into the ambient lamina?
Can one not write down a vector in $\3Q$ describing this organization?

It is proposed that the imaginary $i$ of quantum theory is like the $i$
of  {\sc ac}  circuit theory:
Both represent a contingent  symmetry of  the  environment, an {\em ac} generator in one case and 
a quantum simplicial plexus in the other.
When the generator hunts or the lamina is dislocated,
 the constant $i$ gives way to a dynamical one.
  $\Qi$ 
hopefully also serves as Higgs field, 
inflicting mass on any  gaugeon that does not respect it.

\subsection {\em Acknowledgments}
While I own the errors in this work, I owe important guidance
 to the following among others.
 Von Neumann (1927) made the eye-opening observation that  classical logic,
 namely the {\em Aussagenkalk\"ul},
is a good approximation only for large quantum numbers,
though I first met it in a later paper by him and G. Birkhoff.
\nocite{VONNEUMANN1927}
 David Bohm lectured at Stevens Institute of Technology and Columbia in 1950
 on exterior algebraic simplicial complexes of fermionic vertices, 
 and later wrote them up (1960). 
 Roger Penrose  set me on the spinorial path by  sharing  his ``mops" with me in 1960,
 years before they appeared in print as spin networks, in the same 
 exchange where I
 shared with him the ``unidirectional surface" that became a black hole.
 Richard Feynman (1964) kindly showed me his 
 spinorial space-time quantization, possibly unpublished,
 and raised the space-time Umklapp problem, to be treated elsewhere.
Tenzin Gyatsu (1998) deeply impressed on me the impermanence of natural
 laws in 1998, in the dialogue reported by Arthur Zajonc  (1997).
 \nocite{GYATSU1997, ZAJONC1997}
This sent me back to
 Irving Segal's  (1951) Darwinian theory of the evolution of physical laws toward simplicity,
and his simple quantum space-time.
Still later I found this space-time
 in C. N. Yang (1947), where
I had read it earlier without grasping its significance.

 For more recent helpful and enjoyable discussions I thank James Baugh, Dustin Burns, David Edwards,
 Shlomit Ritz Finkelstein, Andrei Galiautdinov, Dennis Marks, Zbigniew Oziewicz, Heinrich
 Saller, and Sarang Shah.

\end{document}